\documentclass{appolb}%
\usepackage{graphicx}
\usepackage{amsmath}

\def\vec#1{\ensuremath{\mathchoice
                     {\mbox{\boldmath$\displaystyle#1$}}
                     {\mbox{\boldmath$\textstyle#1$}}
                     {\mbox{\boldmath$\scriptstyle#1$}}
                     {\mbox{\boldmath$\scriptscriptstyle#1$}}}}

\begin{document}
%
\title{Relativistic variables for covariant, retardless wave equations
}
\author{M. De Sanctis \footnote{mdesanctis@unal.edu.co}
\address{Universidad Nacional de Colombia, Bogot\'a, Colombia }
\\
}
\maketitle
\begin{abstract}
A reduced  form of the Dirac equation has been previously introduced and studied
in the Center of Mass reference frame. 
In this work we show that this equation 
can be written in a covariant form in a generic reference frame by using specific 
momentum variables.
These variables are also consistent with the retardless form of the interaction 
of the model. 
\end{abstract}
\section{Introduction}
In a previous work  \cite{localred},  we developed a  local reduction of
many-body  relativistic equations 
(more precisely, the Dirac-like equation (DLE) and 
the Mandel\-zweig-Wallace equation  (MW) \cite{mwa})
 for studying the spectroscopy of quark composed systems. 
An accurate calculation of the Charmonium spectrum was performed using a small
 number of free parameters \cite{rednumb}.
In that work a specific form of the regularized vector interaction was used
\cite{chromomds}.\\
In general, for the theoretical formulation of the model we used the Center of Mass reference frame 
(CMRF) where the hadronic bound system is at rest. \\
This choice is perfectly legitimate, in the sense that the internal dynamics of 
the bound system can be studied completely in that frame.\\
However, in order to understand in more detail the relativistic character of the model,
it is  useful to develop its covariant version in a generic reference frame (GRF).
This covariant version of the relativistic equation can be also used to study the 
scattering processes of the hadronic systems.\\
The derivation of the wave equation in covariant form, for a two-body hadronic system, 
represents  the main objective of the present work.
In this context, we show that the retardless and  local character of the interaction 
used in Ref. \cite{localred},
is fully compatible with the relativistic covariance properties of the model.\\
The technique used  in this work to obtain the covariant form 
of the relativistic equation is not completely new: 
similar procedures can be found, for example, 
in Refs. \cite{moshnik,mwb,dspb}.
Here we highlight the specific role of the relativistic variables,
assuming that the total energy of each particle represents the
time component of its four-momentum.
As a consequence, the particles are not on-shell.
Furthermore, the particle energy is considered as a ``fixed" quantity, 
determined in the CMRF by the internal dynamics of the bound system.
Due to this choice, the time component of the momentum transfer
in the CMRF is vanishing, giving a retardless interaction operator.\\
In classical words, each particle of the \textit{bound state} produces
a static  field with which  the other particle interacts.
\vskip 0.5 truecm
\noindent 
For clarity, we point out that the model assumptions introduced above
significantly differ form the relativistic scheme in which one particle
is considered on-shell, 
as, for example,  in the \textit{Relativistic Spectator Formalism}
developed in Refs.
\cite{grossa,grossb,grossc,cst1}.
\vskip 0.5 truecm
\noindent 
Our choice of the relativistic variables and the procedure for obtaining 
a two-body covariant relativistic equation are applied to different 
cases of CMRF relativistic equations.
Finally, we put in covariant form also the reduction operators
introduced in \cite{localred}.
In this way the covariant form of the 
\textit{correlated Dirac wave functions}  is determined.
In this regard, we note that our choice of relativistic variables
is fully consistent with the definition of the reduction operators.

\vskip 1.0 truecm
\noindent
For a thorough description of the CMRF relativistic  model 
and for a comparison
with other relativistic equations, the reader is referred to
Ref. \cite{localred}. 
In the present work we focus our attention on its ``covariantization"
by means of suitable variables.\\
The remainder of the paper is organized as follows.
In Subsection \ref{symbnot} we introduce the symbols and notations used in the work.
In Section \ref{varcor} we discuss the relativistic variables of the model
and analyze their Lorentz transformations.
In particular, in Subsection \ref{basvar} we introduce the basic variables;
in Subsection \ref{lordir} we study the transformations
of the four-vectors and of the Dirac wave functions of the model;
in Subsection \ref{relmg} the momentum variables for a GRF are determined.
In Section \ref{covwe} the covariant form of the wave equation is obtained
focusing the attention on the Dirac-like case; the covariant generalization 
of other forms of the relativistic equation is analyzed 
in Subsection \ref{otherforms}.
The covariant expression of the correlated Dirac wave functions is studied
in Section \ref{covcorr}.
Finally, in Appendix  \ref{detparten},
we briefly discuss 
the method to fix the value of the particle energy.


\subsection{Symbols and Notation}\label{symbnot}
In this work,  the quantities defined as four-vectors will be denoted as in 
the following example: $V=(V^0, \vec V)$.
The Lorentz indices will be written only when strictly necessary and 
in the invariant products; 
for example
$V^\mu U_\mu=V^0 U^0- \vec V \cdot \vec U$.\\
The superscript $c$, not used in \cite{localred}, denotes here
 a four-vector (and also a  wave function)  referred to the CMRF.
No specific symbol is used for the same quantity in a GRF.\\
The lower index $i (j)=1,2$ represents the \textit{particle index}.
The particle index is never summed in this work.\\
For the covariant form of the Dirac equation
we use the gamma matrices of the $i$-th particle $\gamma_i^\mu$ in the standard
representation. For the Hamiltonian form of the Dirac operators, we also introduce
$\beta_i=\gamma_i^0$ and the matrices 
$\gamma_i^0 \gamma_i^\mu= ({\cal I}_i,\vec \alpha_i)$.
The symbol ${\cal I}_i $, that   represents the identity $ 4 \times 4$ matrix
for the $i$-th particle, 
will be omitted when not strictly necessary; 
for example
$V_i^0 {\cal I}_i -\vec V \cdot \vec \alpha_i $
will be written as
$V_i^0 -\vec V \cdot \vec \alpha_i$.
In the same way
$V_\mu \gamma_i^\mu +b {\cal I}_i$
will be written as
$V_\mu \gamma_i^\mu + b$.\\
The Dirac wave fuctions will be represented by the letter $\Psi$; 
the spinorial wave functions by the letter $\Phi$.\\
As customary, throughout the work we use the so-called natural units, 
that is $\hbar=c=1$.\\

\vskip 1.0 truecm

\vskip 1.0 truecm
\section{ The variables of the covariant model. Lorentz and Dirac Boost transformations}\label{varcor}
\subsection{The basic variables of the model}\label{basvar} 
We consider a hadronic \textit{bound system}, composed by two spin $1/2$  particles
(a quark and an antiquark) with masses $m_1$ and $m_2$.\\
We assume that in the CMRF the four-momenta of the two particles are:
\begin{equation}\label{p12cm}
\begin{split}
p_1^{c}= (E_1^c, -\vec p^c),\\
p_2^{c}= (E_2^c, +\vec p^c)~
\end{split}
\end{equation} 
where $E_1^c$ and $E_2^c$ represent the ``fixed", \textit{constant} energy values of the two particles.
This point is studied in more detail in the Appendix \ref{detparten}
where, in eq. (\ref{auxcond}), a standard auxiliary prescription is
recalled to determine their values, shown in eq. (\ref{e12cm}).
The sign  of the internal three-momentum $\vec p^c$ is defined as in 
Refs. \cite{localred,rednumb}.\\
Note that the total four-momentum eigenvalue in the CMRF is
\begin{equation}\label{ptotcm}
P^{c }=p_1^{c }+p_2^{c }= (M , \vec 0) 
\end{equation}
where $M$ represents the mass of the bound system.\\
For the relevant case of two equal mass particles, 
as given in eq. (\ref{e12cmeq}) of  Appendix \ref{detparten},
we have
\begin{equation}\label{eqmp}
E_1^c=E_2^c={\frac M 2}~.
\end{equation}

\vskip 0.5 truecm
\noindent
Also in a generic reference frame (GRF)  the bound system 
is  an eigenstate of
total four-momentum with eigenvalue
\begin{equation}\label{ptot}
P= (E, \vec P),
\end{equation}
where $\vec P$ is the total three-momentum of the bound system and
the total energy $E$ has the standard on-shell expression 
\begin{equation}\label{etot}
E=\sqrt{M^2 +\vec P^2}~.
\end{equation}
In this work, for simplicity, we shall not write explicitly 
the total momentum eigenfunction that will not be used in the calculations.

\subsection{Lorentz and Dirac Boost Trasformations}\label{lordir}
The content of this subsection is completely standard. 
Here, it is reported and applied to our model in order to improve the self-consistency 
of the paper.\\
Using the  definitions of eqs. (\ref{ptot}), (\ref{etot}) 
for the total four-momentum of the system, 
and recalling that the speed of the bound system in a GRF is 
$\vec \beta= \vec P/E$,
we can 
write the Lorentz transformations of  any CMRF four-vector $V^{c }$
to the same four-vector  $V$ in a GRF.
These transformations have the following standard form:

\begin{equation}\label{tldir}
\begin{split}
 V^0={\frac 1 M}\left( E  V^{c~ 0} + \vec P \vec V^c \right), \\
 \vec V= \vec V^c +{\frac {\vec P} { M}}\left(
{\frac {\vec P \vec V^c} {E+M}} +V^{c~0 }\right).
\end{split}
\end{equation}
The inverse Lorentz transformations  are:
\begin{equation}\label{tlinv}
\begin{split}
 V^{c~0}={\frac 1 M}\left( E  V^{0} - \vec P \vec V \right), \\
 \vec V^c= \vec V +{\frac {\vec P} { M}}\left(
{\frac {\vec P \vec V} {E+M}}- V^{0}\right).
\end{split}
\end{equation}
Some words of comment about the first expression of eq. (\ref{tlinv}):
the time component of a four-vector ``seen" in the CMRF is 
a Lorenz invariant quantity.
In consequence, we can rewrite  that expression 
in explicitly invariant form, that is:
\begin{equation}\label{v0inv}
V^{c~0}= { \frac{P_\mu V^\mu}  {M} }~.
\end{equation}
The Dirac Boost operator, for the $i$-th quark, 
has the following standard form:
\begin{equation}\label{dboost}
B_i=F_B\left[ (E+M)
+ \vec \alpha_i \cdot \vec P \right];
\end{equation}
the inverse operator is:
\begin{equation}\label{dbinv}
B_i^{-1}=F_B\left[ (E+M)
 - \vec \alpha_i \cdot \vec P \right],~\\
\end{equation}
with
\begin{equation}\label{factboost}
F_B=\left[2M ( E+M) \right]^{-1/2}~.\\
\end{equation}
We recall that the $B_i$ are used to transform  CMRF Dirac 
wave function
$\Psi^c$ to a GRF Dirac wave function $\Psi$;
furthermore, $\bar \Psi^c$ is transformed with the inverse Boost operators,
that is:  
\begin{equation}\label{psitransform}
\begin{split}
B_1 B_2 \Psi^c= \Psi,~\\
\bar \Psi^c B_1^{-1} B_2^{-1} =\bar \Psi.~\\
\end{split}
\end{equation}
Due to the properties of the Dirac Boost, one has:
\begin{equation}\label{pgamtrans}
B_i V_\mu^c\gamma_i^\mu B_i^{-1}= V_{\mu}\gamma_i^\mu
\end{equation}
where, for a given four-vector $V^c$, $V$ is given by eq. (\ref{tldir}).
In particular, taking 
the unit four-vector
$u^c=(1,\vec 0)$  and consequently
$u=P/M$,
one has:
\begin{equation}\label{gam0trans}
B_i \gamma_i^0 B_i^{-1}= {\frac {P_{\mu} \gamma_i^\mu  } {M}}~.
\end{equation}
%
%


%


\subsection{The  momentum variables in a GRF}\label{relmg}
We can now construct the momentum variables in a GRF.
Starting from eq. (\ref{p12cm}) with the transformations 
of eq. (\ref{tldir})  one can obtain
the four-momenta of the two particles $p_1,~p_2$  in a GRF.\\
However, in order to write the wave equation, it is necessary to introduce, 
in a GRF,  the total and the relative
four-momenta, denoted as $P$ and $p$, respectively. \\ 
The total four-momentum is simply
related to the particle momenta by the standard expression
\begin{equation}\label{ptotp12}
P= p_1+p_2~,
\end{equation}
consistently with the CMRF definition of eq. (\ref{ptotcm}),
 with eq. (\ref{ptot}) and with the Lorentz transformations of eq. (\ref{tldir}).\\
The relative four-momentum can be defined as in Ref. \cite{itz}:
\begin{equation}\label{relfourmom}
p= -\eta_2 p_1
       + \eta_1 p_2~.
\end{equation}
The \textit{constants} $\eta_1$ and $\eta_2$ must satify the condition
\begin{equation}\label{condeta}
\eta_1+\eta_2=1
\end{equation}
and can be \textit{conventionally} chosen as in the nonrelativistic case:
\begin{equation}\label{etai}
\eta_i={\frac {m_i} {m_1+m_2}}~.
\end{equation}
For bound systems  of two equal mass particles, one simply has
$\eta_1=\eta_2=1/2$.\\
From the previous equations one can express the particle momenta $p_i$ 
by means of $P^\mu$ and $p^\mu$ in the following way:
\begin{equation}\label{p1p2}
p_i= p_i(P;p)=\eta_i P- \tau_i p~,
\end{equation}
with $\tau_1=+1$ and $\tau_2=-1$.
In the remainder of the work the four-momenta $p_i$ will be always
considered as functions of the total and relative four-momenta,
as given by the previous equation.

%

\vskip 0.5 truecm
\noindent
For convenience, we introduce  the following definition for the time component
 of the relative four-momentum  in the CMRF:
\begin{equation}\label{deltadef}
p^{c~0}= \Delta~.
\end{equation} 
Eqs. (\ref{p12cm}) and  (\ref{relfourmom}), referred to the CMRF, can be used to 
determine the value of $\Delta$ by means of $E_1^c$ and $E_2^c$.
For the case of equal mass particles (see eq. (\ref{eqmp})) one simply has $\Delta=0$.
In eq. (\ref{deltadif})  of the Appendix \ref{detparten}, we give the explicit value of $\Delta$
for two particles of different mass.
That value is obtained by using the  standard prescription 
of eq. (\ref{auxcond}) 
for determining the particle  energies in the CMRF.\\
In any case, $\Delta$ is  a \textit{fixed} quantity of the bound system;
as  a consequence, we anticipate that the time component of the momentum transfer
in the interaction operator,
is always vanishing in the CMRF, as it will be shown in eq. (\ref{qcmrf}) of the next Section \ref{covwe}. \\
We can write
\begin{equation}\label{prelcmu}
p^{c}= (\Delta, \vec p^c)
\end{equation}
and apply the transformation equations (\ref{tldir})
to determine $p$ in a GRF.\\
For further developments, by means of eq. (\ref{v0inv}),
we can also write:
\begin{equation}\label{deltainv}
\Delta={\frac {P_\mu p^\mu} {M} }
\end{equation}
from which we obtain the expression of $p^0$ in a GRF:
\begin{equation}\label{p0grf}
p^0= {\frac {M \Delta} {E}} + {\frac {\vec P \cdot \vec p} {E}}~.
\end{equation}
In   Subsection \ref{otherforms}
we shall also use explicitly the CMRF particle energies $E_i^c$. 
These \textit{invariant} quantities   can be written taking,
in the CMRF,  the time component of eq. (\ref{p1p2}).
By using the definition of  eq. (\ref{deltadef}),
one obtains
\begin{equation}\label{eic}
E_i^c= \eta_i M-\tau_i \Delta~.
\end{equation}

\vskip 0.5 truecm
\noindent
Finally, in order to determine the invariant relative momentum integration element,
we set $V=p$ and $V^c =p^c$ in
the second relation of eq. (\ref{tldir}) and calculate the Jacobian determinant.
Taking into account that
$p^{c~0}= \Delta$ is a \textit{constant} 
(and, for simplicity, choosing $\vec P$ along a coordinate axis),
one finally finds:
\begin{equation}\label{d3ptransf}
d^3p^c= {\frac M E} d^3p\\
\end{equation}
that represents the covariant integration element in a GRF.

\section{The covariant wave equation in a GRF}\label{covwe}
The objective of this section is to write  in a GRF the  retardless wave equation
of our model.
We discuss here the general procedure, referring, \textit{for definiteness}, to the DLE.
Other forms of equation will be examined 
in the Subsection \ref{otherforms}.
We anticipate that the final result will be written as an integral wave equation.\\
As first step we recall the standard  CMRF wave equation, as it was written in 
the previous work \cite{localred}.  We have
\begin{equation}\label{dleold}
[D_1 O_2+ D_2 O_1 +W ] |\Psi^c>=0~.
 \end{equation}
In the \textit{Hamiltonian form} that was used in \cite{localred}, 
the Dirac operators $D_i$ are:
\begin{equation}\label{oldddef}
D_i=  -(E_i^c - \vec \alpha_i \cdot \vec p_i^c)  + \beta_i m_i 
\end{equation}
and, for the case DLE equation,
\begin{equation}\label{oidef}
O_i^{DLE}={\cal I}_i~.
\end{equation}
\vskip 0.5 truecm
\noindent
We write the same relativistic  equation  as an integral equation
in the momentum space:
\begin{equation}\label{dlehammom}
\begin{split}
\left[~ [ -(E_1^c - \vec \alpha_1 \cdot \vec p_1^c)  + \beta_1 m_1 ] O_2+
        [ -(E_2^c - \vec \alpha_2 \cdot \vec p_2^c)  + \beta_2 m_2 ] O_1 ~
\right]\Psi^c(\vec p^c) \\
+\int d^3p'^c W(\vec q^c) \Psi^c(\vec p'^c) =0~.
\end{split}
\end{equation}
The interaction term has been written in the momentum space as:
\begin{equation}\label{wint}
<\vec p^c | W |\Psi^c>=
\int d^3p'^c W (\vec q^c) \Psi^c(\vec p'^c)
\end{equation}
where $\vec q^c= \vec p^c -\vec p'^c$
represents the three-momentum transfer in the CMRF. 
Note that being $p^{0~c}=\Delta$ a fixed quantity,
the time-component of the momentum tranfer $q^{c~0}$ is always \textit{vanishing} 
\begin{equation}\label{qcmrf}
q^c=(0, \vec q^c)
\end{equation}
\vskip 0.5 truecm
\noindent
that gives rise to  a retardless (or instantaneous)  interaction.
Furthermore, the dependence of the interaction on $\vec q^c$,
corresponds to local form of the interaction in the coordinate space.
This form was used in our previous works \cite{localred,rednumb}. 

\vskip 0.5 truecm
\noindent 
The next step consists in multiplying the DLE of eq.(\ref{dlehammom})
 by $\gamma_1^0 \gamma_2^0$ from the left
in order to write the Dirac operators in the so-called \textit{covariant form}
that is traditionally used to study the Lorentz transformation properties 
in the Dirac theory.
The result is:
\begin{equation}\label{dlecmrf}
\begin{split}
\left [(-p_{1 \mu}^{c}\gamma_1^\mu +m_1)\gamma_2^0 O_2+
       (-p_{2 \mu}^{c}\gamma_2^\mu +m_2)\gamma_1^0 O_1 \right] \Psi^c(\vec p^c)\\
+\int d^3p'^c V(\vec q^c) \Psi^c(\vec p'^c) =0
\end{split}
\end{equation}
where the \textit{invariant} interaction, written in the covariant form, is
\begin{equation}\label{vcov}
V(\vec q^c)=\gamma_1^0 \gamma_2^0 W(\vec q^c)~.
\end{equation}
%

\vskip 0.5 truecm
\noindent
In the last step, recalling that  eq. (\ref{dlecmrf})  is still written in the CMRF,
 we shall replace in that equation the GRF covariant Dirac expressions.\\
In more detail, 
from  eq. (\ref{psitransform}), 
we make the replacement    $\Psi^c= B_1^{-1} B_2^{-1} \Psi$.
%
%
%
Furthermore, we express the particle four-momenta $p_{i}$  by means of 
the total ($P$) and relative ($p$) four-momenta by using
the definition of eq. (\ref{p1p2}).\\
We also premultiply the equation  by $B_1 B_2$ and use eq. (\ref{pgamtrans}) to transform 
the terms with $p_{i \mu}^{c}\gamma_i^\mu $
and eq. (\ref{gam0trans}) to transform
the operators $\gamma_i^0 O_i$
(we are considering here the $O_i$ of  eq. (\ref{oidef}) for the DLE).\\
Finally, the covariant momentum integration is performed by means of eq. (\ref{d3ptransf}).
For convenience, we also multiply by $-1$ the whole equation, obtaining: 
\begin{equation}\label{dlecov}
\begin{split}
\left [( p_{1 \mu}\gamma_1^\mu -m_1) \Omega_2+
       ( p_{2 \mu}\gamma_2^\mu -m_2) \Omega_1 \right] 
			\Psi(p; P)\\
-~{\frac M E}
\int d^3p'~  V( p-p' ) \Psi(p'; P) =0
\end{split}
\end{equation}
where $p -p' $ represents the four-momentum transfer in a GRF
obtained by transforming $q^c$ of eq. (\ref{qcmrf})
by means of eq. (\ref{tldir})
 in standard way.
The covariant operators $\Omega_i$, obtained transforming the $\gamma_i^0$
as explained before,
for the DLE equation, take the form
\begin{equation}\label{omegadle}
\Omega_i^{DLE}={\frac {P_\mu \gamma_i^\mu } {M}} ~.
\end{equation}

\vskip 0.5 truecm
\noindent
Eq. (\ref{dlecov}) represents the covariant, retardless equation of our model.\\
In order to discuss a different procedure for its derivation,
 we can write it as an equation for invariant matrix-elements,
multiplying by $ \bar \Psi(p;P)$ and performing the covariant integration:
\begin{equation}\label{dleinvm}
\begin{split}
 {\frac M E}  \int d^3p  \bar \Psi(p;P)
\left [( p_{1 \mu}\gamma_1^\mu -m_1) \Omega_2+
       ( p_{2 \mu}\gamma_2^\mu -m_2) \Omega_1\right] 
			\Psi(p; P)\\
-\left({\frac M E}\right)^2
 \int d^3p_b  \int d^3p_a  
\bar \Psi(p_b;P)
 V( p_b-p_a ) \Psi(p_a; P) =0~.
\end{split}
\end{equation}
One factor $M/E$ is obviously redundant and can be cancelled; it has been 
written explicitly in the previous equation in order to highlight the covariant character
of the integrations.

\vskip 0.5 truecm
\noindent
We shall now derive eq. (\ref{dleinvm}) in a slightly different way, in order to analyze
in more detail the properties of the relative four-momentum  and 
the corresponding covariant integration procedure  of our model.\\
 The relative four-momentum $p$ (and $p_a,~p_b$) of eq. (\ref{dleinvm}) is
a \textit{constrained} quantity, because in the CMRF its time component is
\textit{fixed}, as given  in eqs.  (\ref{deltadef} ) and (\ref{prelcmu}).\\
We consider \textit{only for this derivation}, that is for the next 
eq. (\ref{fourdimeq}),
an \textit{unconstrained} relative four-momentum $p$ and the unconstrained
particle four-momenta $p_i$  that are expressed by means of $p$ and $P$ 
with the same relation of eq. (\ref{p1p2}).
But, to recover eq. (\ref{dleinvm}),  we have to introduce a \textit{constraint function} that
will be discussed below.\\
We can write an explicitly four-dimensional matrix-element equation in the form

\begin{equation}\label{fourdimeq}
\begin{split}
\int d^4p ~\theta(p;P) \bar \Psi(p;P)
\left [( p_{1 \mu}\gamma_1^\mu -m_1) \Omega_2+
       ( p_{2 \mu}\gamma_2^\mu -m_2) \Omega_1 \right] 
			\Psi(p; P)\\
-\int d^4 p_b ~\theta(p_b;P)  \int d^4p_a~ \theta(p_a;P) 
\bar \Psi(p_b;P)
 V( p_b-p_a ) \Psi(p_a; P) =0
\end{split}
\end{equation}
where $\theta(p;P)$ represents the  \textit{constraint function}.\\
In our model,
the covariant constraint function  is written by means of the Dirac delta
 function in the form:
\begin{equation}\label{constrfunc}
\theta(p;P)= \delta \left( {\frac 1 M} P_\mu p^\mu -\Delta\right)
= {\frac M E} \delta \left[ p^0 - \left(
 {\frac {M \Delta} {E}} +{\frac {\vec P \cdot\vec p} {E} } \right)     \right]~.
 \end{equation}
We find that, using this constraint function, eq. (\ref{dleinvm}) is immediately recovered 
with the same value of 
$p^0$,  given in eq. (\ref{p0grf}), for a GRF.\\
The previous derivation can be also used as  a starting point to study
 different forms of constraint function $\theta(p;P)$
for reproducing the physical spectroscopy of the hadronic systems.

\subsection{Other forms of the wave equation}\label{otherforms}
We study here the covariant form of other  wave equations that are obtained
by replacing the operators of eq. (\ref{oidef}) with other expressions.\\
We have shown that, in the case of  the DLE, taking the 
 $O_i^{DLE}$ of eq. (\ref{oidef}) for the CMRF operators in the Hamiltonian form,
we obtain the $\Omega_i^{DLE}$ of eq. (\ref{omegadle}) for the
GRF operators in the covariant form.

\vskip 0.5 truecm
\noindent
In our previous works we have also considered the MW equation.
In that model the operators $O_i^{MW}$ 
have the form
\begin{equation}\label{oimw}
O_i^{MW}={\frac {\vec \alpha_i \cdot \vec p_i^c + \beta_i m_i}
           {\sqrt{m_i^2 +({{{\vec p_i}^c})^2}} }}~.
\end{equation}
These operators were denoted as $S_i$ in the work \cite{localred} because they
represent the energy sign of the free particle in the CMRF.\\
In order to find the corresponding $\Omega_i^{MW}$ ,
one has to multiply  the operators  of eq. (\ref{oimw}) 
  by $\gamma_i^0$ from the left
 and then determine their  covariant GRF expression.
With standard handling one obtains:
\begin{equation}\label{omegamw}
\Omega_i^{MW}=
 {\frac {\tau_i \cdot (p_\mu - {\frac \Delta M} P_\mu)\gamma_i^\mu + m_i}
           {\sqrt{m_i^2 +\Delta^2 - p_\nu p^\nu  } } }
\end{equation}
with $\tau_i$ defined just after eq. (\ref{p1p2}).
Note that the previous expression takes a simple form  when $\Delta=0$,
corresponding to the case of equal mass particles. 

\vskip 0.5 truecm
\noindent
Other forms for the operators $O_i$ and $\Omega_i$ can be studied.
As in work \cite{rednumb} we have analyzed numerically some specific expressions
for the charmonium spectrum. 
In particular, we obtained very similar results as those published 
in Ref. \cite{rednumb} by using the following operators. In the  ``Model A", we used
\begin{equation}\label{oimoda}
O_i^{A}=  {\frac {E_i^c -\vec \alpha_i \cdot \vec p_i^c} {E_i^c}}~.
\end{equation} 
The corresponding covariant GRF operators are:
\begin{equation}\label{omegamoda}
\Omega_i^{A}= {\frac { p_{i\mu} \gamma_i^\mu} {E_i^c}}~.
\end{equation}

\vskip 0.5 truecm
\noindent
In the ``Model B" we have replaced in eqs. (\ref{oimw}) and (\ref{omegamw})
the particle masses $m_i$ with the CMRF energies $E_i^c$. 
We have
\begin{equation}\label{oimodb}
O_i^{B}={\frac {\vec \alpha_i \cdot \vec p_i^c + \beta_i E_i^c}
 {\sqrt{(E_i^{c})^2  +({{{\vec p_i}^c})^2}} }}~
\end{equation}
and
\begin{equation}\label{omegamodb}
\Omega_i^{B}=
 {\frac {\tau_i \cdot (p_\mu - {\frac \Delta M} P_\mu)\gamma_i^\mu + E_i^c}
           {\sqrt{(E_i^c)^2 +\Delta^2 - p_\nu p^\nu  } }} ~. 
\end{equation}
In the previous eqs. (\ref{oimoda}) - (\ref{omegamodb}), 
the CMRF particle energies $E_i^c$  are given by  eq. (\ref{eic}).

\section{ The covariant form for the correlated	Dirac wave function}\label{covcorr}
In this section we study the problem of constructing the correlated Dirac wave function
in a GRF consistently with the covariant form of the model.
In the work \cite{localred} the correlated wave function was determined  only in the CMRF.
Now we have to boost this wave function  to a GRF.
In more detail, we can write:
\begin{equation}\label{boostpsicorr}
\Psi_{corr}(p;P)= B_1 B_2 \Psi_{corr}^c(\vec p^c)
\end{equation}
\vskip 0.5 truecm
\noindent
where $\Psi_{corr}^c(\vec p^c)$ represents the CMRF correlated Dirac wave function
introduced in  Ref.  \cite{localred}.
Furthermore, the Dirac Boost operators $B_1$ and $B_2$ of eq. (\ref{dboost}) are standardly used here.\\
In principle, the previous expression of $ \Psi_{corr}(p;P)$ would be sufficient 
to study the bound state problem  in any GRF.
However, we prefer to express $ \Psi_{corr}(p;P)$ by means of explicitly covariant
quantities.\\

\vskip 0.5 truecm
\noindent
To this aim, one has to recall, in the first place,
that   the relative momentum $\vec p^c$ can be written, with the Lorentz tranformations 
(\ref{tlinv}),
in terms of the GRF momenta $p$ and $P$;  
these quantities are the  arguments 
of $\Psi_{corr}$ in the \textit{l.h.s.}  of eq. (\ref{boostpsicorr}).\\
Then, we elaborate the  boosted  correlated wave function.
\vskip 0.5 truecm
\noindent
In the  CMRF, the correlated  Dirac wave function 
was obtained  by means of the reduction  operator $K_i^c$, 
of the form:
\begin{equation}\label{defk}
K_i^c= 
\begin{pmatrix} 1 \\ 
                {\frac {\vec \sigma_i \cdot \vec p_i^c} {m_i+E_i^c}  }
\end{pmatrix}~.
\end{equation}
In more detail, for a two-body system, one has to apply the reduction operators of
 the two particles
\begin{equation}\label{psiccorr}
\Psi_{corr}^c= K_1^c K_2^c \Phi^c(\vec p^c)
\end{equation}
where $\Phi^c(\vec p^c)$ represents the spinorial wave function.
The numerical normalization constant is calculated apart, as in eq. (38) of 
Ref. \cite{localred} 
and omitted in the following.

\vskip 0.5 truecm
\noindent
For the present procedure of ``covariantization" we preliminarly introduce 
the following operator
\begin{equation}\label{lambdadef}
\Lambda_i= 
\begin{pmatrix} 1 \\
0
\end{pmatrix}
\end{equation}
that, applied to $\Phi^c$, simply constructs a Dirac spinor 
with vanishing lower components.\\
Then, by means of $\Lambda_i$, we can express, with standard calculations, 
the  operator $K_i^c$   in the following form
\begin{equation}\label{kicnew}
 K_i^c=F_i^c
(p_{i \mu}^c \gamma_i ^\mu+ m_i) \Lambda_i~,
\end{equation}
with the invariant factor
\begin{equation}\label{fic}
F_i^c=(E_i^c + m_i)^{-1}~.
\end{equation}
The expression of eq. (\ref{kicnew}) is equivalent to eq. (\ref{defk}) but
is more suitable to be transformed into a covariant expression.\\
We can  apply the Boost operator of eq. (\ref{dboost})
 to the $K_i^c$ of eq. (\ref{kicnew}) 
in order to determine the reduction operator $K_i$ in a GRF.
We can write
\begin{equation}\label{kigrf}
K_i=B_i K_i^c=F_i^c B_i (p_{i \mu}^c \gamma_i ^\mu+ m_i)B_i^{-1}B_i \Lambda_i~.
\end{equation}
With standard calculations, one finds:
\begin{equation}\label{bilambdai}
B_i \Lambda_i = F_B (P_\mu \gamma_i^\mu + M ) \Lambda_i~.
\end{equation}
Furthermore, the factor in parenthesis of eq. (\ref{kigrf})
can be transformed by means of eq. (\ref{pgamtrans}).
The result is:
\begin{equation}\label{kif}
K_i= F_B F_i^c (p_{i \mu} \gamma_i ^\mu+ m_i)(P_\mu \gamma_i^\mu + M ) \Lambda_i~.
\end{equation}
By using the reduction operators of the two particles, one obtains
the correlated wave function in a GRF, in the form:
\begin{equation}\label{psicorfin}
\begin{split}
\Psi_{corr}(p;P)=K_1 K_2 \Phi^c= 
(F_B)^2 F_1^c F_2^c  (p_{1 \mu} \gamma_1 ^\mu+ m_1)(p_{2 \mu} \gamma_2 ^\mu+ m_2)\\
\times (P_\mu \gamma_1^\mu + M )(P_\mu \gamma_2^\mu + M )\Lambda_1 \Lambda_2 \Phi^c~.
\end{split}
\end{equation}
The correlated Dirac adjoint wave function
can be easily constructed by means of standard handling of the Dirac matrices.
One has  to use also the following property of the operator $\Lambda_i$ 
\begin{equation}\label{lambdabar}
\Lambda_i^\dag \gamma_i^0= \Lambda_i^\dag~.
\end{equation}
The result is
\begin{equation}\label{psibarcorfin}
\begin{split}
\bar \Psi_{corr}(p;P)= \Psi_{corr}^\dag(p;P) \gamma_1^0 \gamma_2^0=
\Phi^{c\dag} \Lambda_1^\dag \Lambda_2^\dag
(P_\mu \gamma_1^\mu + M )(P_\mu \gamma_2^\mu + M )\\
\times (p_{1 \mu} \gamma_1 ^\mu+ m_1)(p_{2 \mu} \gamma_2 ^\mu+ m_2)
(F_B)^2 F_1^c F_2^c~.
\end{split}
\end{equation}
Eqs. (\ref{psicorfin}) and (\ref{psibarcorfin}) complete the development of this
section by using explicitly covariant  operators for the Dirac wave function.
The result shows 
 that the use of the correlated wave function in a GRF is consistent 
with the whole covariant model. 
Finally, we note that
the factor 
$(F_B)^2 F_1^c F_2^c$, being a \textit{constant}, is not relevant for the covariant
wave equation.
Furthermore, in principle, the  covariant integral wave equation
of eq. (\ref{dlecov}), with the Dirac correlated wave function
of eq. (\ref{psicorfin}),
could be solved 
in a GRF, determining \textit{directly} the corresponding spinorial function.

\newpage
\vskip 0.5 truecm
\centerline{{\bf Acknowledgements}}
The author thanks the group of  ``Gesti\'on de Recursos de Computo Cient\'ifico,
Laboratorio de Biolog\'ia Computacional,
Facultad de Ciencias - Universidad Nacional de Colombia"
for the access to the computation facilities that were used to perform the numerical calculations that have been necessary to develop the model discussed 
and generalized in this work.

\appendix
\section{Determination of the particle energy}\label{detparten}
%
Considering eq. (\ref{p12cm}), we observe that
the CMRF energy values  $E_1^c$ and $E_2^c$ are usually determined by means 
of the  auxiliary prescription \cite{mwa,mwb}:
\begin{equation}\label{auxcond}
(E_1^c)^2 - (E_2^c)^2 = m_1^2 -m_2^2
\end{equation}
that is related, in general, to the asintotic properties of the relativistic
free Hamiltonian of the two-body system.\\
The CMRF energies $E_i^c$ are easily found
 by using eqs. (\ref{ptotcm}) and (\ref{auxcond}). One has
\begin{equation}\label{e12cm}
\begin{split}
E_1^c= {\frac 1 2} (M+ {\frac {m_1^2- m_2^2} {M}}),~\\
E_2^c= {\frac 1 2} (M- {\frac {m_1^2- m_2^2} {M}})~.
\end{split}
\end{equation}
The validity of eqs. (\ref{auxcond}) and (\ref{e12cm}) 
for a bound system
in the general case  of $m_1 \neq m_2$ should be carefully verified.
However, for the specific case $m_1=m_2$, that is physically very relevant 
for the study of the $q~\bar q$ mesons, one simply has
\begin{equation}\label{e12cmeq}
 E_1^c=E_2^c={\frac M  2} ~.
\end{equation}
For completeness, starting from eq. (\ref{e12cm}), 
one can detemine $\Delta$ of eq. (\ref{deltadef}).
By using eq. (\ref{relfourmom}) for the time component of the relative momentum 
in the CMRF and the definition of  the $\eta_i$ 
given in eq. (\ref{etai}) of
Subsection \ref{relmg},
one obtains:
\begin{equation}\label{deltadif}
\Delta= {\frac M 2} {\frac {m_1-m_2} {m_1+m_2}} -
{\frac 1 2} {\frac {m_1^2 -m_2^2} {M}}~.
\end{equation}

\vskip 0.5 truecm
\noindent
For the case of two equal mass particles,
the standard result of eq. (\ref{e12cmeq}),
 can be also obtained in a different way:
 one can use the
symmetry properties of the system and 
consider $E_1^c$ and $E_2^c$ as \textit{mean values} of the
corresponding Hamiltonian operators.\\
Due to the relevance of this argument in the context of this work, 
we summarize its derivation.\\
We introduce the CMRF Hamiltonian, 
schematically written in the form:
\begin{equation}\label{hcmrf}
H^c = H_{1~free}^c + H_{2~free}^c +W 
\end{equation}
where the first two terms represent the free Hamiltonian of the two particles
and $W$ is the interaction term.
The corresponding eigenvalue equation is:
\begin{equation} \label{eigeqcm}
H^c |\Psi^c>= M|\Psi^c>~.
\end{equation}
Given that the two particles have the same mass, the interaction is symmetric
with respect to particle interchange:
\begin{equation}
P_{12} W P_{12} =W 
\end{equation}
where $P_{12} $ represents the particle interchange operator.\\
In consequence, the eigenstates have definite symmetry:
\begin{equation}\label{statesym}
P_{12} |\Psi^c>= (-1)^\sigma |\Psi^c>
\end{equation}
with $\sigma=+1$ for \textit{symmetric} and $\sigma=-1$ for \textit{antisymmetric}
states, respectively.\\
We can define the interacting Hamiltonian operator for each particle in the form:
\begin{equation}\label{hinti}
H_{i}^c= H_{i~free}^c +{\frac 1 2} W
\end{equation}
with $i=1,2$.
In this way, one can immediately sum the  Hamiltonians
of the two particles to obtain
the expression of eq. (\ref{hcmrf})
\begin{equation}\label{hcsum}
H^c=H_1^c+H_2^c~.
\end{equation} 
Furthermore, the standard interchange
property
\begin{equation}\label{hicsym}
P_{ij} H_{i}^c P_{ij} = H_{j}^c
 \end {equation}
is automatically satisfied.\\
By using eqs. (\ref{eigeqcm}), (\ref{statesym}), (\ref{hcsum}) and (\ref{hicsym}),
 one finds
\begin{equation}
<\Psi^c| H_{1}^c |\Psi^c>= <\Psi^c| H_{2}^c |\Psi^c> ={\frac M 2}~.
\end{equation}
 Further investigation is needed to study the case of two particles with 
different masses.\\

\newpage


\end{document}